\documentclass[aps,prl,twocolumn,superscriptaddress,floatfix,showpacs]{revtex4}

\usepackage{float}
\usepackage{placeins}
\usepackage{amssymb,amsmath}
\usepackage{graphicx}
\usepackage[colorlinks=true,linkcolor=blue,citecolor=blue,urlcolor=blue]{hyperref}

\begin{document}

\title{Presence of a fluctuation related peak but absence of vortex-lattice melting in Nb$_3$Sn in high resolution specific-heat measurements}

\author{M.\ Reibelt}
\email[]{reibelt@physik.uzh.ch}
\affiliation{Physik-Institut University of Zurich, Winterthurerstrasse 190, CH-8057 Zurich, Switzerland}
\author{N.\ Toyota}
\affiliation{Physics Department, Graduate School of Science, Tohoku University, 980-8571 Sendai, Japan}
\date{\today}

\begin{abstract}
The melting of the magnetic vortex lattice has been observed in high-T$_c$ superconductors in many experiments by different groups and is regarded as confirmed. To date, only one group claims to have observed the vortex-lattice melting in the low-$T_c$ superconductor Nb$_3$Sn in specific-heat measurements. We measured the same Nb$_3$Sn single crystal with a differential-thermal analysis method. We report on the absence of any sign of vortex-lattice melting in our data and discuss the possible reasons for this discrepancy. In addition we confirm the observation of a small peak-like anomaly near the transition to superconductivity which is likely related to thermal fluctuations.
\end{abstract}


\maketitle

\section{INTRODUCTION}
Close to the transition to superconductivity fluctuations manifest in several physical quantities like the specific heat, the magnetization, and the resistivity. The Ginzburg number is a measure of the importance of thermal effects for the respective superconductor. The larger the Ginzburg number the more important thermal fluctuations become for the calculation of the physical properties of the superconductor. In layered superconductors extreme physical properties such as large anisotropy, short coherence length, and high transition temperatures combine to enhance thermal fluctuations. A quantity for the estimation of the width of the critical region, where fluctuations are of importance, is the Ginzburg temperature $\tau_G$, which is related to the Ginzburg number $G_i$ by $\tau_G = T_{c0} G_i$. According to Thouless \emph{et al.}~\cite {Thouless1960}, the fluctuation specific heat of clean bulk superconductors is not expected to be observable until $(T-T_{c0})/T_{c0} \approx 10^{-11}$. The situation improves with dirty samples and thin films, since they have increased Ginzburg numbers. In the specific heat these fluctuations manifest as a peak-like anomaly peak close to the transition to superconductivity which we will call the \emph{fluctuation peak} in the following. The fluctuation peak is a rarely observed phenomenon in low-$T_c$ superconductors. So far, to the best of our knowledge, only a few observations of the fluctuation peak in low-$T_c$ superconductors are known in the literature. Besides the observation in the recent works of Lortz \emph{et al.}~\cite{Lortz2006September,Lortz2007April}, a similar peak in the heat capacity below $T_c(H)$ was first observed in $1967$ by Barnes and Hake \cite{Barnes1967} for a type-II superconductor in the extreme dirty limit, although they were unable to study the behavior in the critical region because of additional sample broadening of the transition. Effects of fluctuations have also been observed in the specific heat of extremely dirty films by Zally and Mochel \cite{Zally1971December}. Later in $1975$, an enhanced heat capacity above and also below $T_c(H)$ was also observed in measurements on Nb by Farrant and Gough \cite{Farrant1975}; their sample was in the clean limit, a very pure crystal with negligible broadening due to impurities. The high resolution and data-point density of the DTA method used by us benefitted us in the investigation of the suspected fluctuation peak in Nb$_3$Sn. In this paper we present high resolution specific-heat data on a homogeneous single crystal of Nb$_3$Sn. We confirm the observation of the fluctuation peak close to $T_c(H)$ as observed in an earlier work by Lortz \emph{et al.}~\cite{Lortz2006September}.\\
\indent In the so-called \emph{peak effect region} close to $T_c$ the rise of the critical current at the onset of the peak effect has been attributed to an abrupt softening of the shear modulus of the vortex lattice as $H$ approaches the upper critical field $H_{c2}(T)$ \cite{Pippard1969,Larkin1979,Brandt1977,Brandt1977a,Brandt1977b,Brandt1977c,Brandt1993September}, which may has its origin in thermal fluctuations. When the vortex lattice becomes less rigid, the vortices can bend and adjust better to the pinning sites. In line with the collective-pinning description of Larkin and Ovchinnikov \cite{Larkin1970,Larkin1973,Larkin1979} the peak effect points to a transition of the vortex lattice from an ordered phase to a disordered phase. Whether this transition is a thermodynamic phase transition is still a matter of dispute to date. In case it is a true thermodynamic phase transition one expects to find an anomaly in the specific heat and the equilibrium magnetization. The associated metastability of an underlying first-order vortex-lattice melting transition has been proposed as the origin of the peak effect. One assumes that the melting transition is hindered by the strong pinning in the peak-effect region. In order to be able to observe the proposed melting transitions one can try to equilibrate the vortex lattice by the application of a small magnetic so-called \emph{vortex-shaking field} $h_{ac}$ as was done by Willemin \emph{et al.}~\cite{Willemin1998September,Willemin1998November}. According to theories \cite{Brandt1986November,Brandt1994April,Brandt1996March,Brandt1996August,Mikitik2000September,Mikitik2001August,Brandt2002July,Brandt2004,Brandt2004a,Brandt2004b,Brandt2007}, transversal and also longitudinal vortex shaking with a small oscillating magnetic field $h_{ac}$ can cause magnetic vortices to "walk" through a superconductor, which leads to a reduction of non-equilibrium current distributions and thereby to an annealing of the vortex lattice. This technique was applied, for example, for resistivity measurements \cite{Cape1968February}, torque magnetometry \cite{Willemin1998September,Willemin1998November,Weyeneth2009}, local magnetization measurements \cite{Avraham2001May}, and also for specific-heat investigations \cite{Lortz2006September}.
Lortz \emph{et al.}~\cite{Lortz2006September} applied such a shaking field parallel to the main magnetic field ($h_{ac} \parallel H$) to a Nb$_3$Sn single crystal and they claim to have observed a vortex-lattice melting transition in the peak-effect region in high resolution specific-heat data. We also applied a shaking field to the same Nb$_3$Sn sample during some of our high resolution specific-heat measurements, but with a different field configuration was different. However, according to our interpretation we did not observe a vortex-lattice melting transition, in contrast to the work by Lortz \emph{et al.}~\cite{Lortz2006September}.

\section{EXPERIMENT}
The Nb$_3$Sn single crystal (mass $m \approx 11.9\, $mg, thickness $d \approx 0.4\, $mm, and cross section $A \approx 0.44\, $mm$^2$) used for this study was characterized by Toyota \emph{et al.}~\cite{Toyota1988September} and was further investigated by Lortz \emph{et al.}~\cite{Lortz2006September,Lortz2007March,Lortz2007April} in calorimetric measurements and in resistivity measurements in our group \cite{Reibelt2010March}. The crystal exhibits a pronounced peak effect near $H_{c2}$ \cite{Lortz2007March,Reibelt2010March}, and its transition to superconductivity in zero magnetic field as determined by a resistivity measurement occurs at $T_{c} \approx 18.0\, $K.\\
\indent The specific heat was measured with a home-made differential-thermal analysis calorimeter, which achieves a very high sensitivity and data-point density \cite{Schilling2007March,Reibelt2008}. For some specific-heat measurements we applied a small shaking field $h_{ac}$ perpendicular to the main magnetic field $H$ continuously during the measurement. The shaking field was oriented perpendicular to the longest dimension of the sample, that is perpendicular to the critical currents inside the sample, a so-called \emph{transversal vortex-shaking configuration} ($h_{ac} \perp j_c$ and $h_{ac} \perp H$) \cite{Brandt2002July,Brandt2004}. We varied the amplitude of the shaking field to maximal $\mu_0h_{ac} \approx 1.3\, $mT and the frequency varied between $5$ and $100\, $Hz. The vortex lattice preparation prior to the measurement was a field cooled (FC) procedure. For most measurements the FC procedure was very slow and took several hours. The measuring cell was thermally decoupled from the rest of the insert very well, heat exchange took place only via its nylon wire suspension. For a few measurements we used a cooling clamp which decreased the cooling time during the FC procedure to about $15$ minutes.

\section{RESULTS AND DISCUSSION}
\subsection{A. Fluctuation peak}
\begin{figure}
\includegraphics[width=75mm,totalheight=200mm,keepaspectratio]{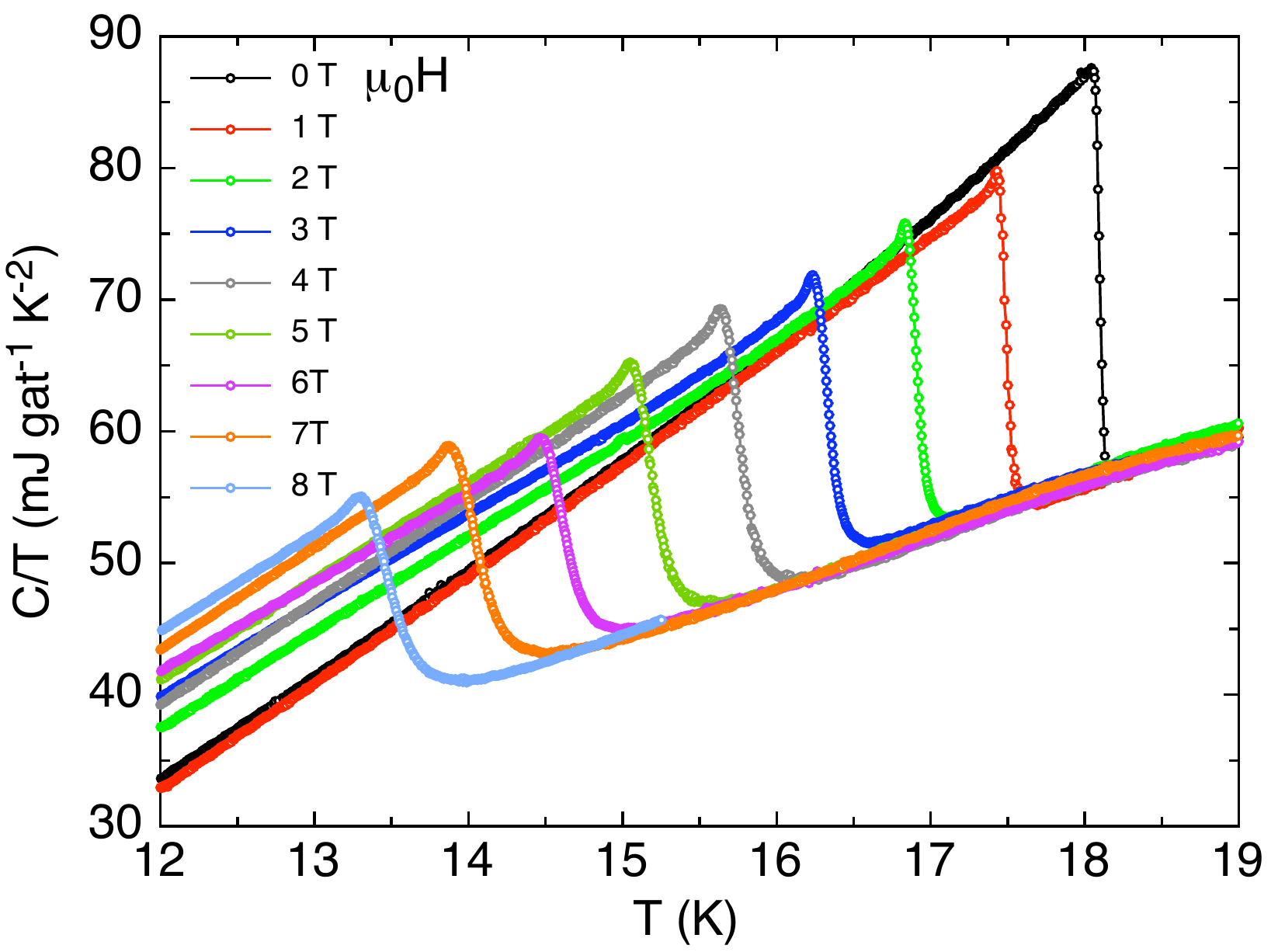}
\caption{Specific heat of a Nb$_3$Sn single crystal for different fixed magnetic fields from $0\, $T to $8\, $T. Here $1\, $gat (gram-atom) is $1/4\, $mole.\label{fig.PRB_2012_all_fields_pdf}}
\end{figure}
\begin{figure}
\includegraphics[width=75mm,totalheight=200mm,keepaspectratio]{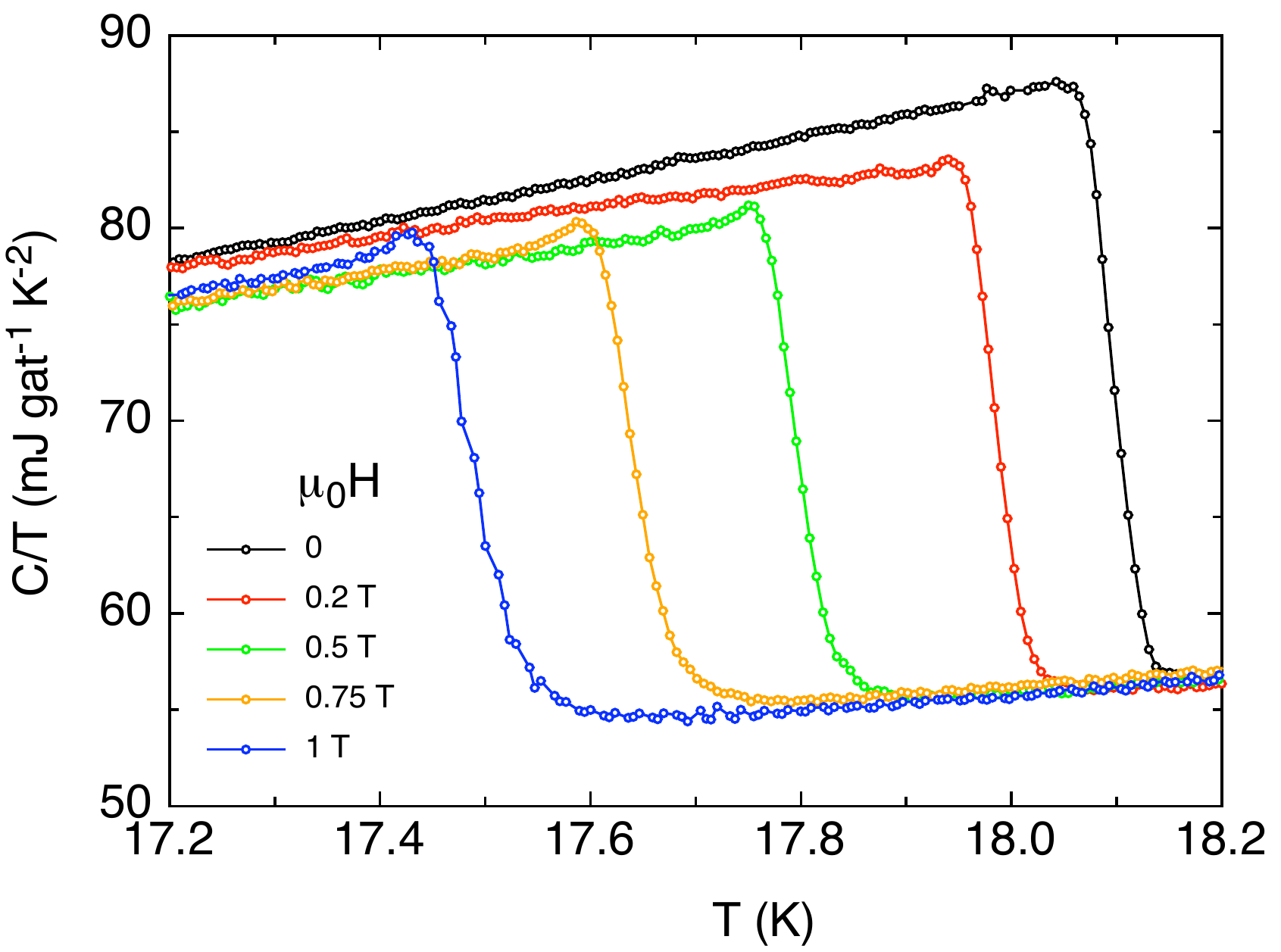}
\caption{Development of the fluctuation peak in the specific heat of Nb$_3$Sn at low magnetic fields.\label{fig.PRB_2012_fluctuation_peak_development_pdf}}
\end{figure}

In Fig.~\ref{fig.PRB_2012_all_fields_pdf} we plotted the temperature dependence of the specific heat at different fixed magnetic fields, note the high data-point density. A peak-like anomaly near $T_c(H)$ is discernible for all shown nonzero magnetic fields, the fluctuation peak. This fluctuation peak was also seen by Lortz \emph{et al.}~\cite{Lortz2006September,Lortz2007April} in the same crystal. With increasing magnetic field, the transition widths at $T_c$ increases and the fluctuation peak broadens. The existence of the fluctuation peak and the increase of the transition width can be attributed to the presence of critical fluctuations of the superconductivity order-parameter and a one-dimensional ($1$D) character in the presence of a magnetic field \cite{Eilenberger1967,Thouless1975April}. Since the nature of the fluctuation is changed rather drastically when a magnetic field is applied, one expects that the critical region will be broadened \cite{Lee1972April}, which in turn leads to a broadening of the fluctuation peak. Figure \ref{fig.PRB_2012_fluctuation_peak_development_pdf} shows the development of the fluctuation peak at low magnetic fields. It can be discerned already at $\mu_0H=0.2\, $T and its magnitude increases with increasing magnetic field until it saturates at about $\mu_0H=1\, $T where its size reaches $\sim 5\%$ of the total jump of the specific heat at $T_c(H)$. The Ginzburg temperature $\tau_G = 0.5\, k^2_B\, T^3_c(0)/(H^2_c(0)\, \xi^3_0)^2$ determines the temperature range around $T_c$ where the contribution of fluctuations to the specific heat are of the same order of magnitude as the mean-field jump at $T_c$. Using $T_c \approx 18\, $K, $H_c(0) = 5200\, $Oe and $\xi_0 = 30\, ${\AA} \cite{Guritanu2004November}, we obtain $\tau_G \approx 10^{-4}\, $K. Contributions of a few percent of the jump might therefore be observable in a range of $10^{-2}\, $K around $T_c$. As can be seen from Fig.~\ref{fig.PRB_2012_fluctuation_peak_development_pdf}, the fluctuation peak near $T_c(H)$ develops over a temperature range of several $10^{-2}\, $K.\\
\FloatBarrier
\FloatBarrier
\indent According to Lortz \emph{et al.}~\cite{Lortz2007April} for their data the width of the transition to superconductivity and also the width of the fluctuation peak in Nb$_3$Sn scaled according to the $3$D-LLL scaling model for $\mu_0 H \geq 4\, $T and to the $3$D-XY scaling model for smaller magnetic fields. The problem which arises here is the fact that according to the $3$D-XY scaling model, the fluctuation peak should be most pronounced for zero magnetic field. Due to the divergence of the coherence length $\xi$ at $T_c$, the peak should also diverge in an ideal case, but finite size effects due to sample dimensions, sample inhomogeneities lead to a broadening of the peak; in addition the normal conducting cores of the vortices in the presence of a magnetic field lead to a broadening of the peak with increasing magnetic field. In the high-$T_c$ superconductors YBa$_2$Cu$_3$O$_{7-\delta}$ \cite{Lortz2003November} and NdBa$_2$Cu$_3$O$_7$ \cite{Plackowski2005} one has observed this diverging behavior for zero magnetic field. Therefore within the $3$D-XY model it is hard to understand, why in Nb$_3$Sn the peak is absent in zero magnetic field and develops only with increasing magnetic field. However, in the picture of lowest Landau levels (LLL) one may understand the observations. A magnetic field confines the quasiparticles to low Landau levels, thereby reducing the dimensionality of the fluctuations. For very low magnetic fields $\mu_0 H \leq 0.2\, $T, the confinement is not strong enough and no fluctuation peak is formed in the specific heat. With increasing magnetic field the confinement to low-lying Landau Levels gets stronger and as a result only one degree of freedom remains, namely that along the $z$ direction. The fluctuation specific heat is then proportional to the field and becomes one-dimensional in nature, diverging within the mean field theory, like $|T/T_{c}(H)-1|^{-3/2}$ as compared with a $|T/T_{c0}-1|^{-1/2}$ divergence in the absence of a magnetic field. This results in a substantial enhancement of the specific heat close to the transition temperature \cite{Lee1972April} and may explain the observed development of a fluctuation peak in the Nb$_3$Sn specific-heat data above $0.2\, $T.\\
\FloatBarrier
\begin{figure}
\includegraphics[width=75mm,totalheight=200mm,keepaspectratio]{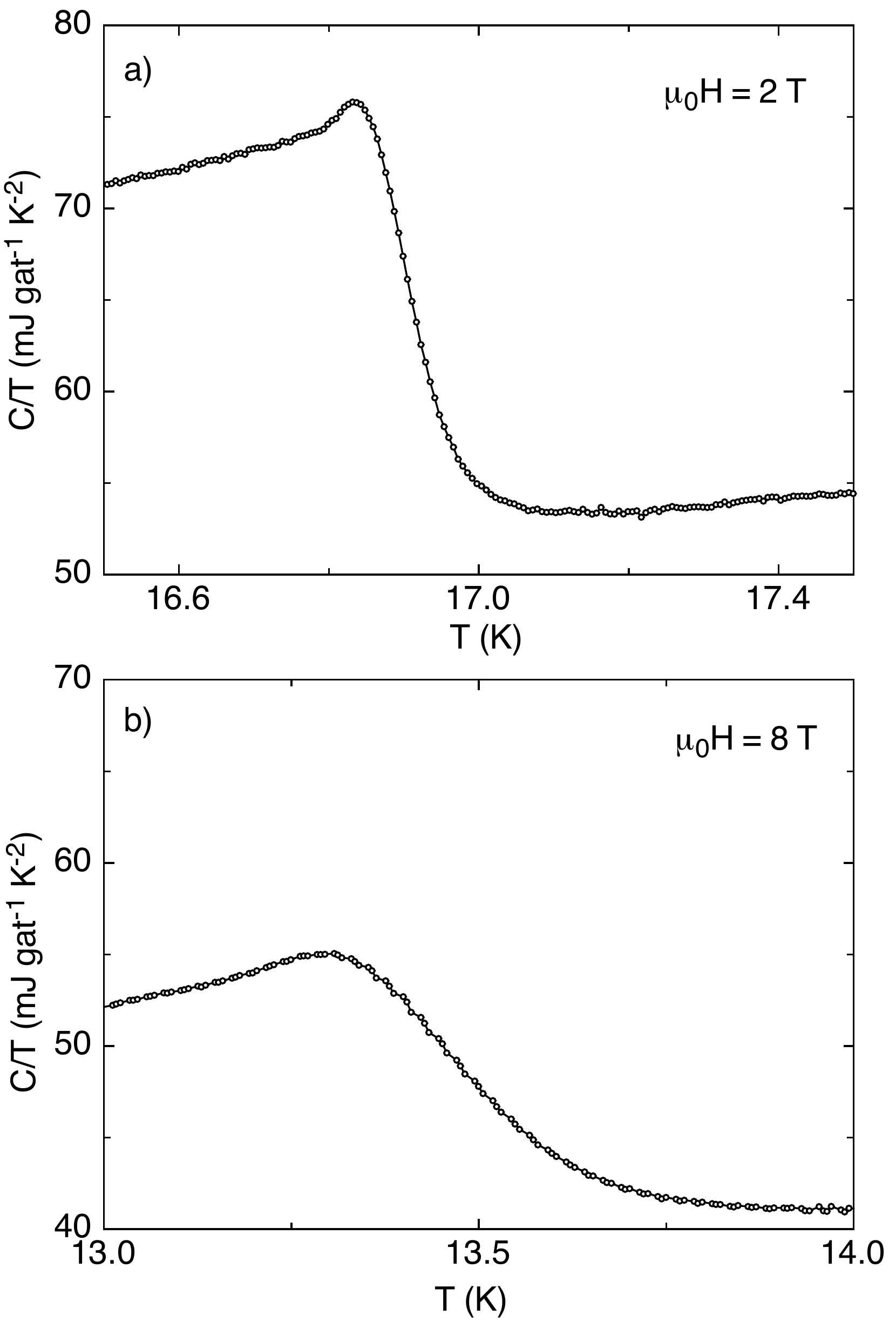}
\caption{Broadening of the fluctuation peak in Nb$_3$Sn: \textbf{a)} Specific heat at $\mu_0H=2\, $T. \textbf{b)} Specific heat at $\mu_0H=8\, $T.\label{fig.PRB_2012_2T_vs_8T_pdf}}
\end{figure}
\FloatBarrier
\indent Besides the initial increase in magnitude of the fluctuation peak, one can also observe a broadening with increasing magnetic field. This is expected, if one keeps in mind that also $\tau_G$ increases in a magnetic field due to a reduction of the effective dimensionality arising from the confinement of the excitations to a few low Landau orbits \cite{Lee1972April}. In order to illustrate the increase of the broadening of the fluctuation peak with increasing magnetic field we compare in Fig.~\ref{fig.PRB_2012_2T_vs_8T_pdf} a measurements at $\mu_0H=2\, $T with a measurement at $\mu_0H=8\, $T. Note that the axes of Figs.~\ref{fig.PRB_2012_2T_vs_8T_pdf}a and \ref{fig.PRB_2012_2T_vs_8T_pdf}b are scaled to the same size.

\subsection{B. Absence of vortex-lattice melting}
\begin{figure}
\includegraphics[width=75mm,totalheight=200mm,keepaspectratio]{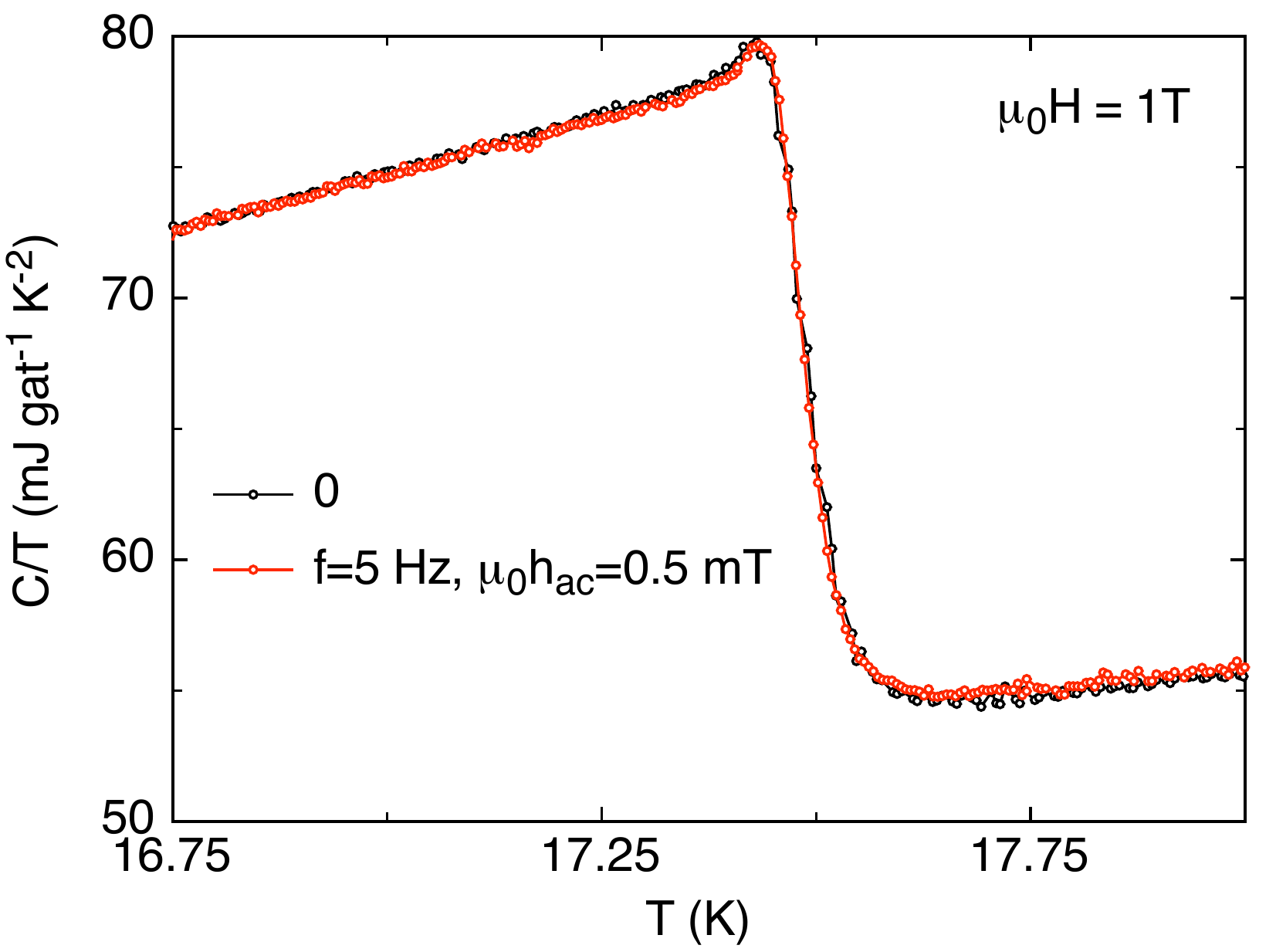}
\caption{Specific heat of Nb$_3$Sn at $\mu_0H=1\, $T. A small shaking field with $h_{ac} \approx 0.5\, $mT and $f = 5\, $Hz was applied perpendicular to the main magnetic field $H$ for the red curve.\label{fig.PRB_2012_1T_5Hz_pdf}}
\end{figure}

\begin{figure}
\includegraphics[width=75mm,totalheight=200mm,keepaspectratio]{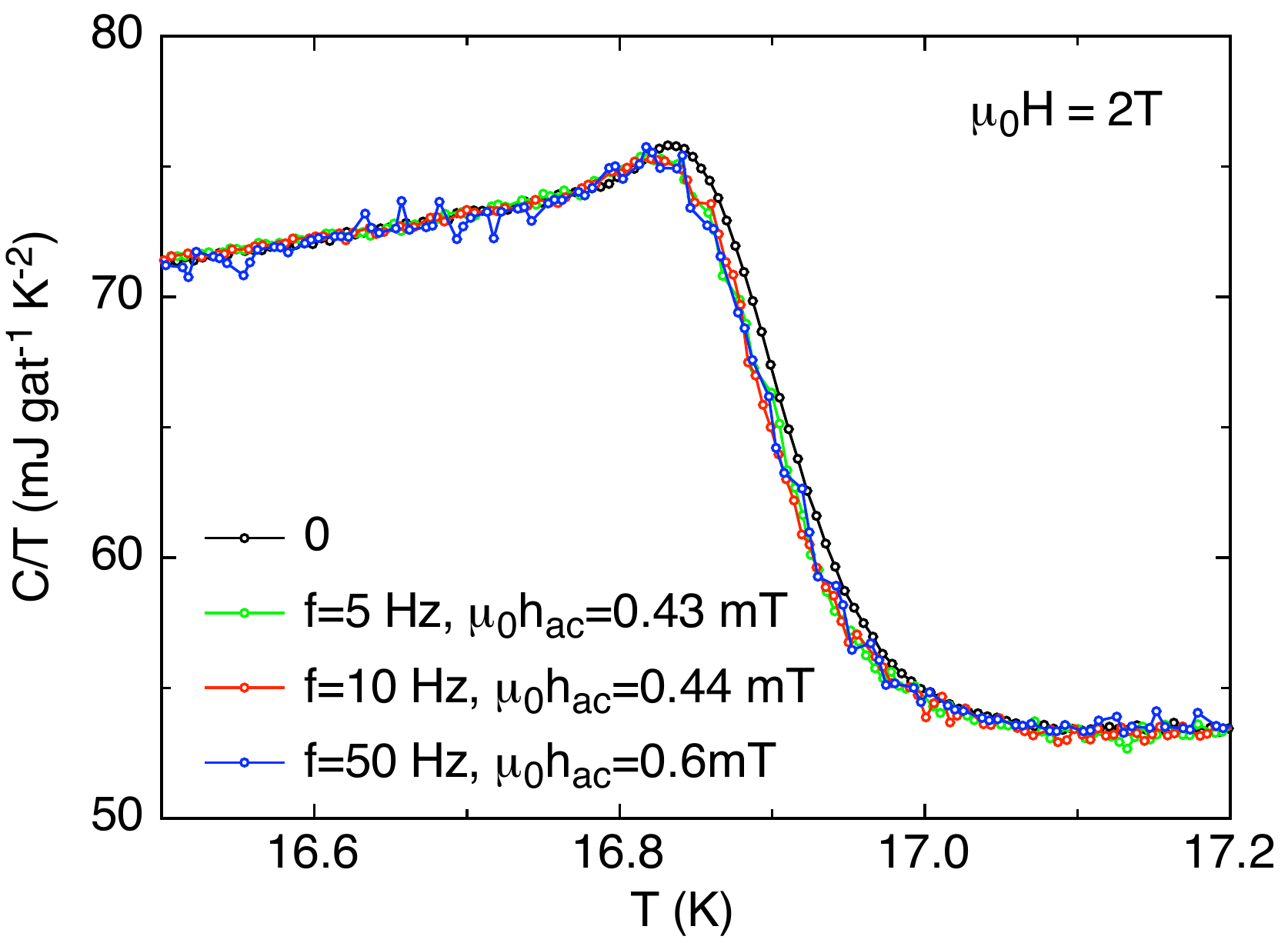}
\caption{Specific heat of Nb$_3$Sn at $\mu_0H=2\, $T. A small shaking field with different amplitudes $h_{ac}$ and frequencies $f$ was applied perpendicular to the main magnetic field $H$.\label{fig.PRB_2012_2T_pdf}}
\end{figure}

\begin{figure}
\includegraphics[width=75mm,totalheight=200mm,keepaspectratio]{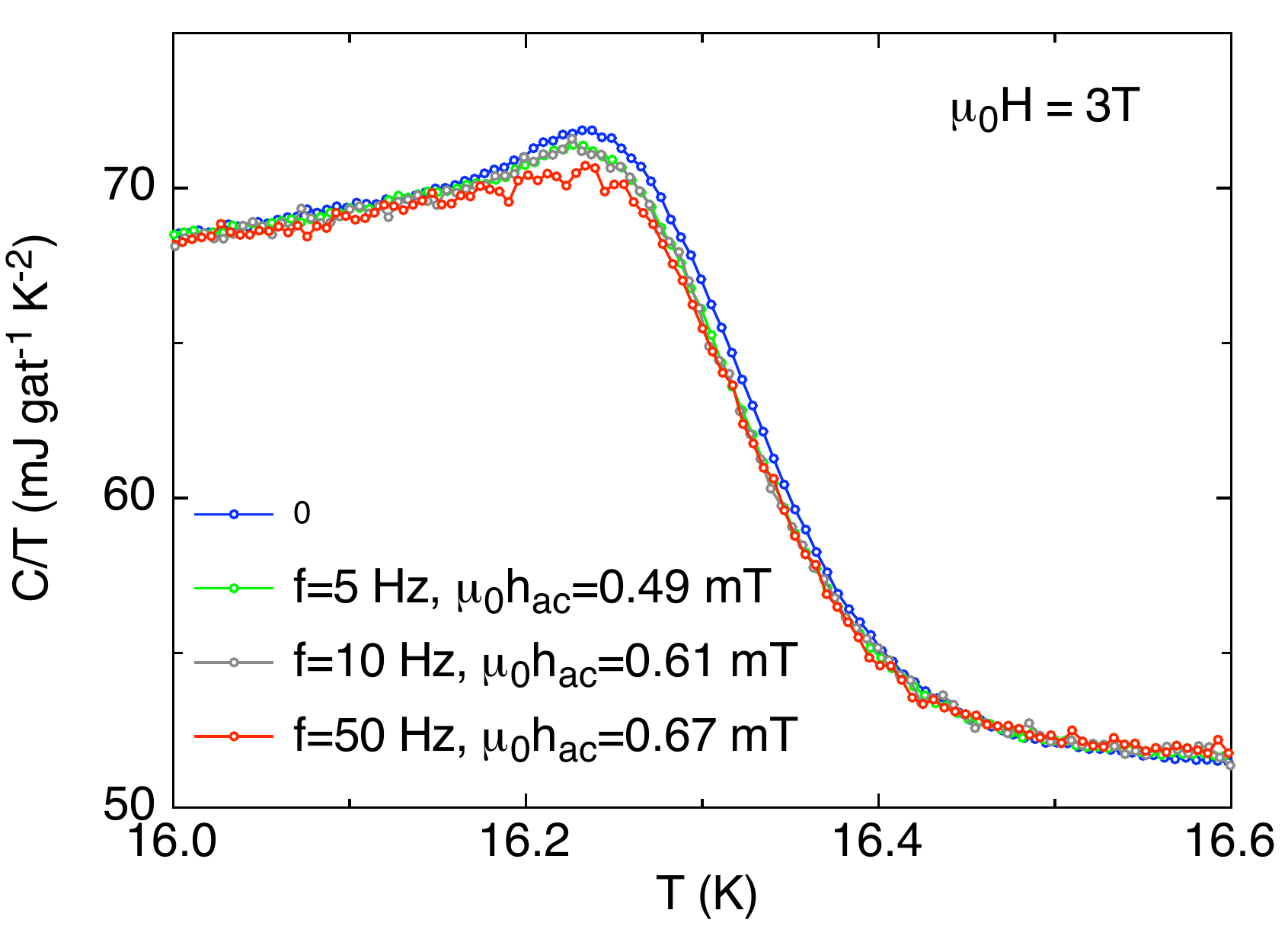}
\caption{Specific heat of Nb$_3$Sn at $\mu_0H=3\, $T. A small shaking field with different amplitudes $h_{ac}$ and frequencies $f$ was applied perpendicular to the main magnetic field $H$.\label{fig.PRB_2012_3T_pdf}}
\end{figure}

\begin{figure}
\includegraphics[width=75mm,totalheight=200mm,keepaspectratio]{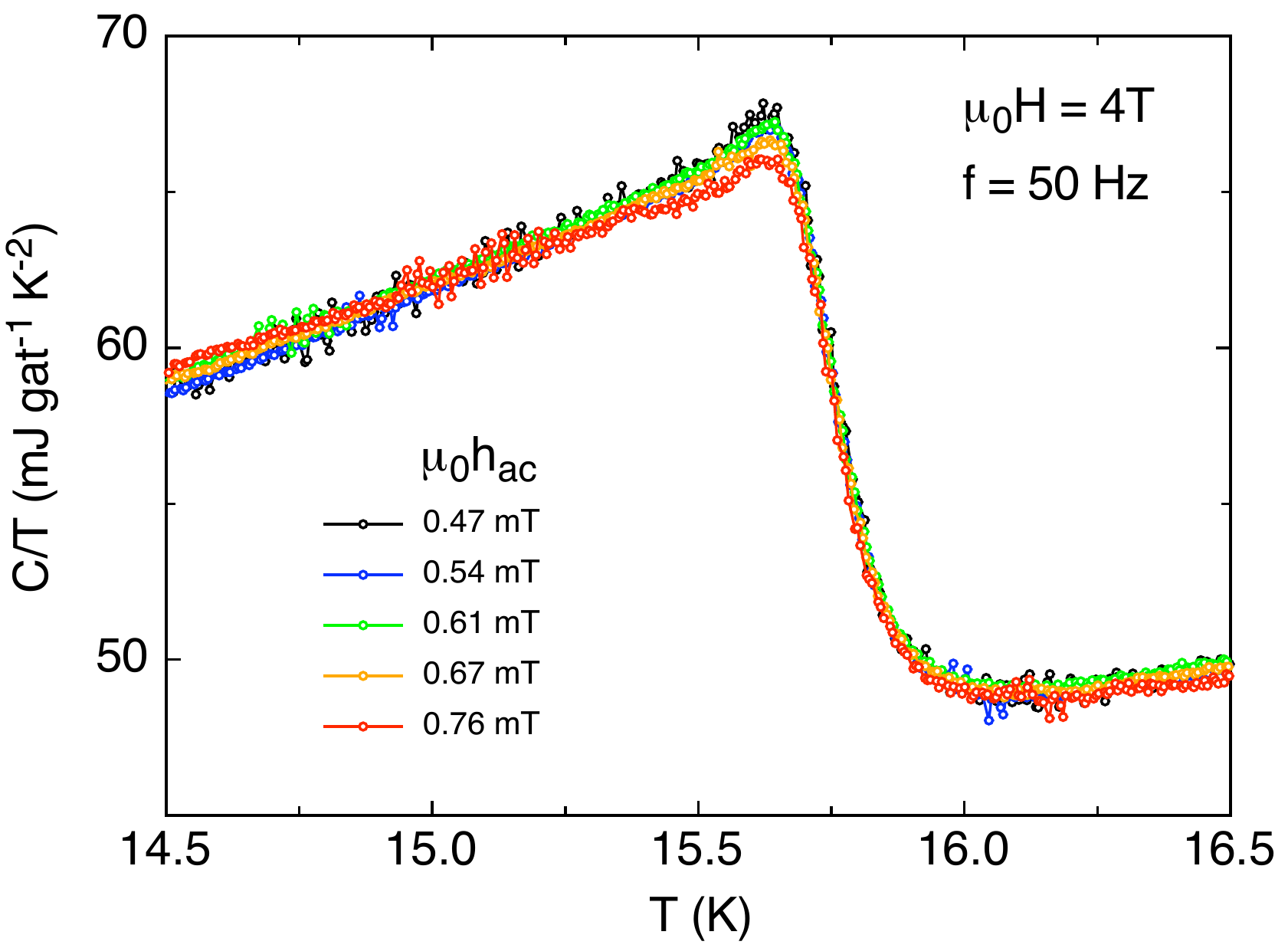}
\caption{Specific heat of Nb$_3$Sn at $\mu_0H=4\, $T. A small shaking field at frequency $f=50\, $Hz with different amplitudes $h_{ac}$ was applied perpendicular to the main magnetic field $H$.\label{fig.PRB_2012_4T_50Hz_pdf}}
\end{figure}

\begin{figure}
\includegraphics[width=75mm,totalheight=200mm,keepaspectratio]{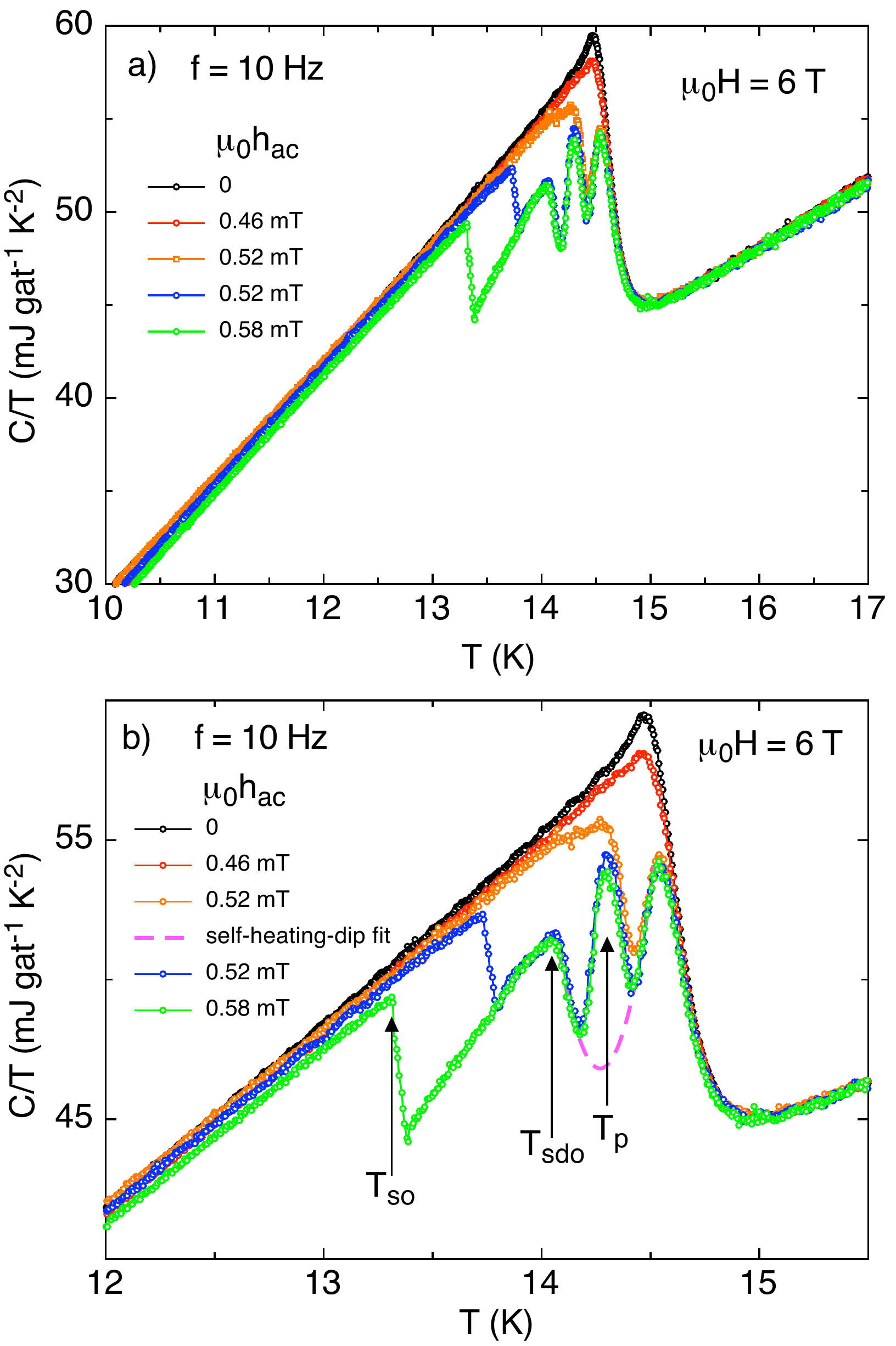}
\caption{\textbf{a)} Distorted specific heat of Nb$_3$Sn at $\mu_0H=6\, $T for different shaking amplitudes $h_{ac}$ at fixed frequency $f = 10\, $Hz. \textbf{b)} Closeup of the region near $T_c$. \label{fig.PRB_2012_6T_10Hz_both_pdf}}
\end{figure}

We next applied a small shaking field at different amplitudes $h_{ac}$ and frequencies $f$ perpendicular to the main magnetic field $H$ continuously during the specific-heat measurements. For low magnetic fields (Figs.~\ref{fig.PRB_2012_1T_5Hz_pdf} and \ref{fig.PRB_2012_2T_pdf}), no peak-like deviation from the data without a shaking field can be noticed in our data.
\FloatBarrier
\indent According to their interpretation, Lortz \emph{et al.}~\cite{Lortz2006September} observed the first sign of vortex-lattice melting at $\mu_0H=3\, $T and they investigated and observed the feature up to $\mu_0H=6\, $T (see also their magnetic phase diagram in Fig.~$6$ of Ref.~\cite{Lortz2007March}). Their observed sharp peak-like feature increased in magnitude with increasing magnetic field as one would expect it for a true vortex-lattice melting transition. In Fig.~\ref{fig.PRB_2012_3T_pdf} and Fig.~\ref{fig.PRB_2012_4T_50Hz_pdf} we present our data at $\mu_0H=3\, $T and $\mu_0H=4\, $T for amplitudes up to $h_{ac} = 0.76\, $mT and frequencies up to $f= 50\, $Hz. We did not note any sign of a sharp peak which may be connected to a latent heat of a first-order melting transition. However, Lortz \emph{et al.}~\cite{Lortz2006September} used a higher amplitude of about $h_{ac} \approx 1\, $mT and a higher frequency of $f=1\, $kHz. Our shaking coil was not able to reach these parameters. However, Lortz \emph{et al.}~\cite{Lortz2006September} applied the shaking field not perpendicular but parallel to the main magnetic field.
\FloatBarrier
\FloatBarrier
\noindent According to Brandt and Mikitik, application of a small shaking field perpendicular (transversal vortex-shaking \cite{Brandt2002July,Brandt2004}) or parallel (longitudinal vortex-shaking \cite{Mikitik2003March}) to the critical current inside the sample is able to cause the critical currents and the irreversible magnetic moment of the sample to relax completely, however application of a small shaking field parallel to the main magnetic field $H$ and perpendicular to the critical currents inside the sample plane is only able to do so if a transport current is applied to the sample simultaneously \cite{Mikitik2001August}. The latter configuration was successfully used by us in resistivity measurements \cite{Reibelt2010March}. Lortz \emph{et al.}~\cite{Lortz2006September} applied a small shaking field $h_{ac} \approx 1\, $mT parallel to the main magnetic field $H$ but they did not apply the necessary transport current to the sample. Therefore, from the current theoretical point of view it is questionable whether their shaking field was able to relax the irreversible currents inside the sample as claimed by Lortz \emph{et al.}~\cite{Lortz2006September}. However, Lortz \emph{et al.}~\cite{Lortz2006September} placed the sample not inside their shaking coil but above it, therefore only the weak external stray field of the shaking coil reached the sample. The external stray field of a shaking coil was not homogeneous and not all its components were parallel to the main magnetic field, some of its field components were perpendicular to the main magnetic field and inside the sample plain. These perpendicular components may have been able to conduct transversal and longitudinal vortex-shaking inside the sample \cite{Brandt2004,Mikitik2003March}. According to Brandt \emph{et al.}~\cite{Brandt1986November,Brandt1994April,Brandt1996March,Brandt1996August,Mikitik2000September,Mikitik2001August,Brandt2002July,Brandt2004,Brandt2004a,Brandt2004b,Brandt2007} there exists a threshold amplitude below which no movement of the vortices is generated. In order to be able to cause the vortices to \emph{walk}, the shaking amplitude has to fulfill the condition $h_{ac} \geq J_c/2$, where $J_c$ is the critical sheet current density, which is the integration of the critical current density $j_c$ over the thickness $d$ of the sample. $J_c$ is proportional to the pinning strength. However, it is questionable whether Lortz \emph{et al.}~\cite{Lortz2006September} were able to reach this necessary threshold shaking-field amplitude $h_{ac} \geq J_c/2$ to cause the "walking" of the vortices \cite{Brandt2002July}. The field components of their shaking field inside the plane of the sample are most likely smaller than the amplitude of the shaking field which was applied by us in the transversal shaking configuration. Regarding the lower frequency used by us, Brandt \emph{et al.}~\cite{Brandt2002July} mention only a threshold amplitude but not a threshold frequency for the transversal shaking configuration. Therefore we assume that the measurements conducted by us should have been able to reveal the vortex-lattice melting transition, if present, even at frequencies $f<1\, $kHz.\\
\indent The presence of the peak effect near the upper critical field in this Nb$_3$Sn sample has been observed by Lortz \emph{et al.}~\cite{Lortz2006September} and was confirmed in our recent work \cite{Reibelt2010March}. Lortz \emph{et al.}~\cite{Lortz2007March} have shown that the peak effect coincides in their magnetic phase diagram with their claimed observation of a vortex-lattice melting transition. However, the presence of the peak effect can complicate the interpretation of data. From the description of the conduction of the experiment in Ref.~\cite{Lortz2006September} it appears that Lortz \emph{et al.}~continuously applied a shaking field during the measurement and did not turn it off while taking the data at each data point. Therefore it might be possible that the sample got heated up due to vortex motion. In the following we will call the heating up of the sample which is not due to the cell heater but due to vortex motion inside the sample itself as \emph{self heating}. Directly at the peak effect the vortex lattice becomes stronger pinned which leads to an increased critical current. In this region of increased pinning the self heating of the sample due to vortex motion should decrease. Depending on the background subtraction procedure and possible adjustments to gain absolute values, this decrease in self heating might have been misinterpreted as a peak-like increase in the specific heat.\\
\begin{figure}
\includegraphics[width=75mm,totalheight=200mm,keepaspectratio]{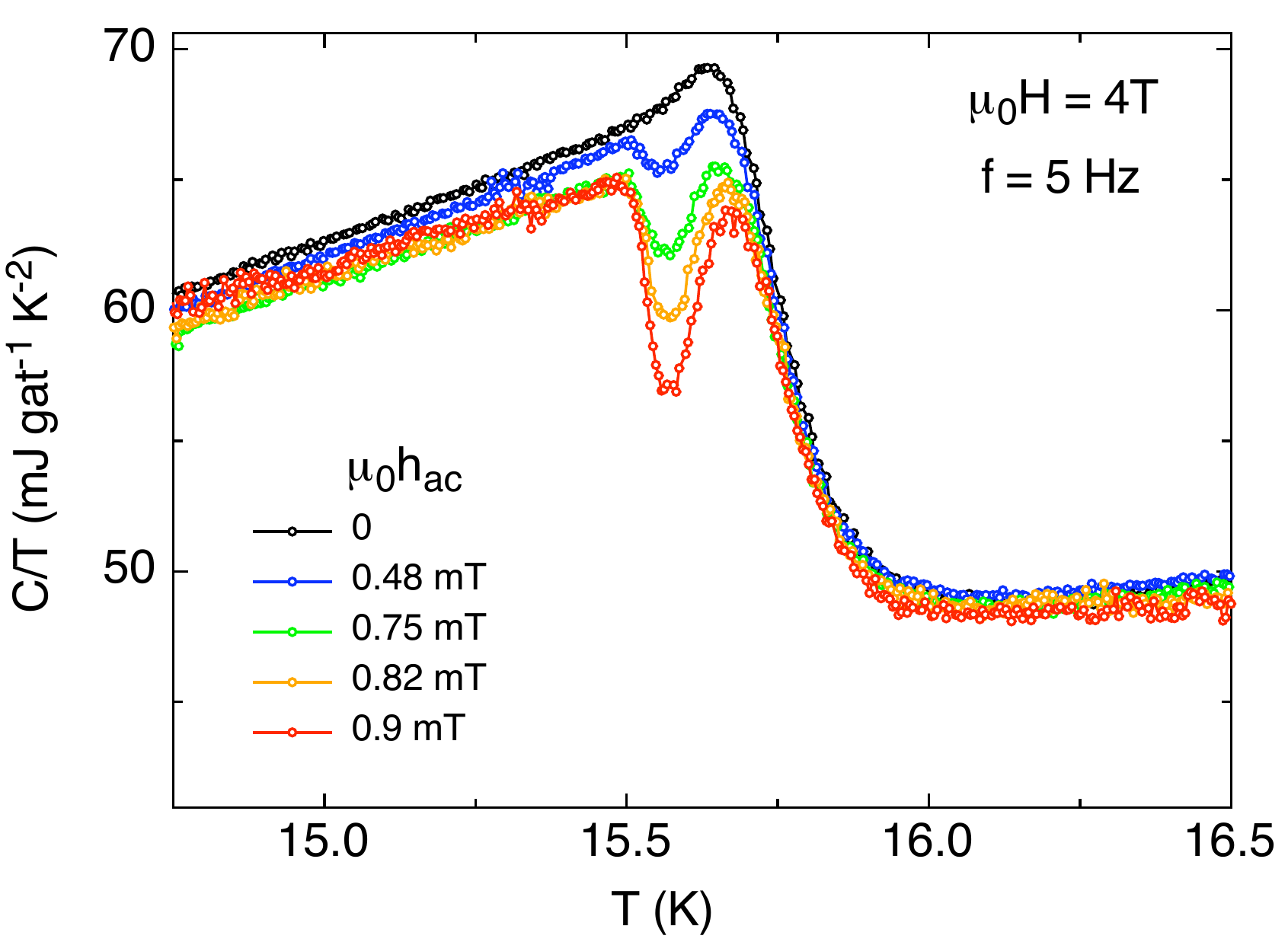}
\caption{Distorted specific heat of Nb$_3$Sn at $\mu_0H=4\, $T, for different shaking amplitudes $h_{ac}$ at fixed frequency $f=5\, $Hz.\label{fig.PRB_2012_4T_5Hz_pdf}}
\end{figure}
\indent In Fig.~\ref{fig.PRB_2012_4T_5Hz_pdf} we set the shaking frequency to $f=5\, $Hz at $\mu_0H=4\, $T. Close to $T_c$ self heating occurs which increases with increasing amplitude of the shaking field $h_{ac}$. The surplus heat created inside the sample causes the DTA calorimeter to underestimate the specific heat of the sample which leads to the formation of a dip-like pattern in the data, which we will call \emph{self-heating dip} in the following. Therefore, for a nonzero shaking field the curves in Fig.~\ref{fig.PRB_2012_4T_5Hz_pdf} do not represent the specific heat in the regions where self heating occurs. However, these data of a distorted specific heat are still of value since a sharp first-order phase transition should still be discernible as a sharp peak since the self-heating causes merely a smooth distortion. However, no superimposed peak can be discerned in our data which we plotted in Fig.~\ref{fig.PRB_2012_4T_5Hz_pdf}.\\
\begin{figure}
\includegraphics[width=75mm,totalheight=200mm,keepaspectratio]{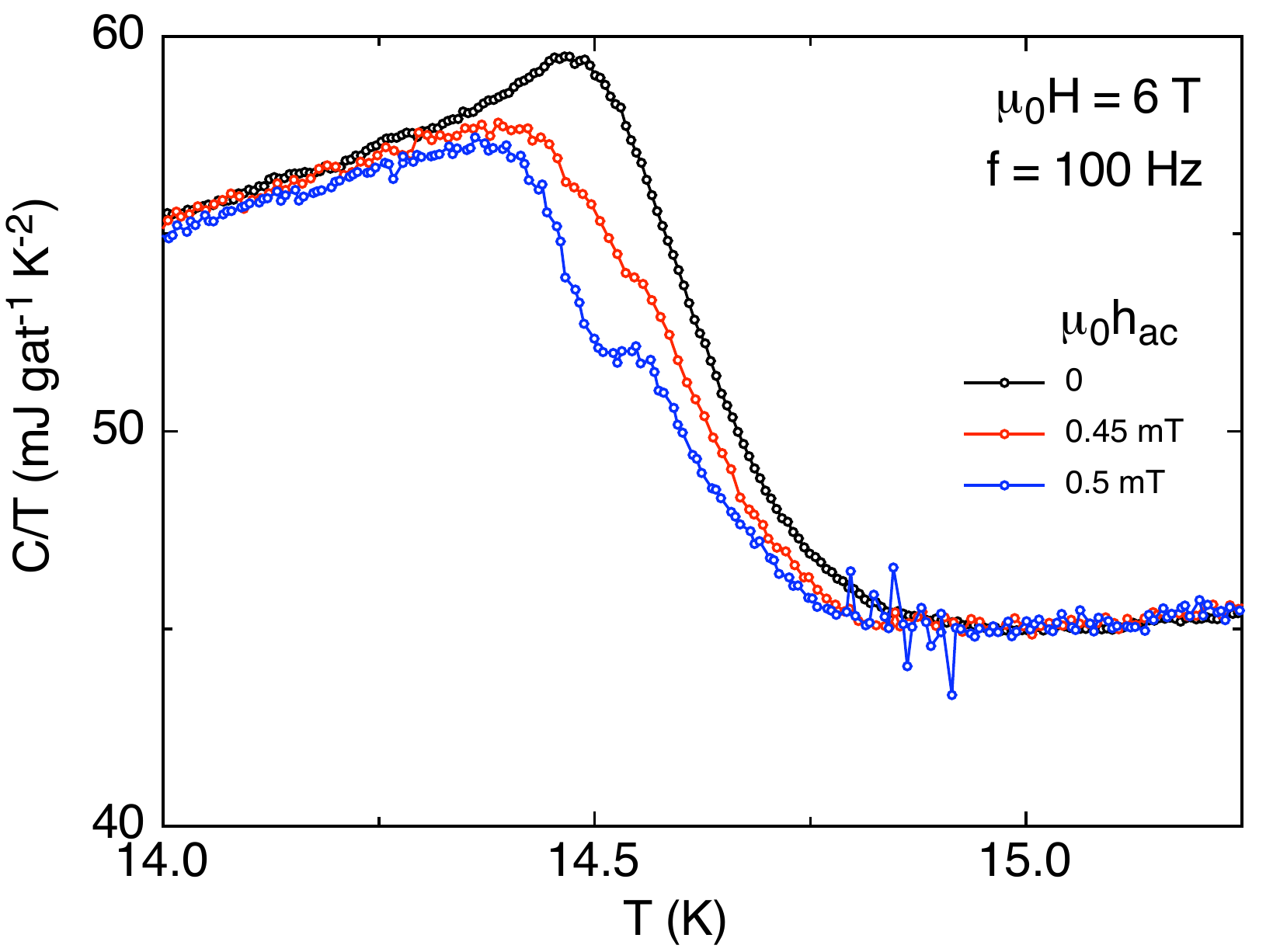}
\caption{Distorted specific heat of Nb$_3$Sn at $\mu_0H=6\, $T. A small shaking field with frequency $f=100\, $Hz and different amplitudes $h_{ac}$ was applied perpendicular to the main magnetic field $H$. Self heating of the sample sets in for $h_{ac}>0$ which distorts the specific heat in a region close to $T_c$.\label{fig.PRB_2012_6T_100Hz_pdf}}
\end{figure}
\begin{figure}
\includegraphics[width=75mm,totalheight=200mm,keepaspectratio]{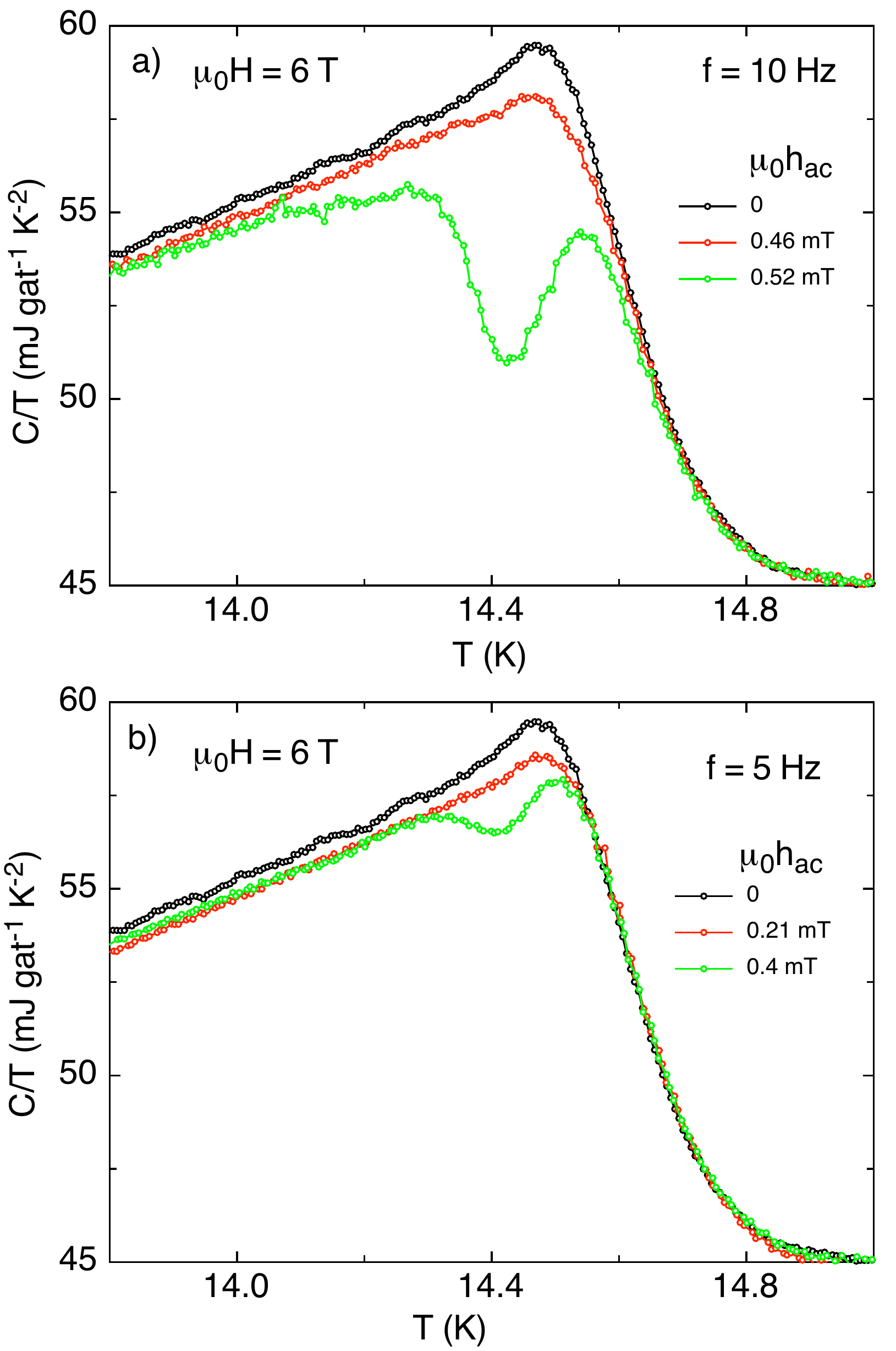}
\caption{Distorted specific heat of Nb$_3$Sn at $\mu_0H=6\, $T. A small shaking field was applied perpendicular to the main magnetic field $H$ at different amplitudes $h_{ac}$. \textbf{a)} $f=10\, $Hz; \textbf{b)} $f=5\, $Hz.\label{fig.PRB_2012_6T_10Hz_pdf}}
\end{figure}
\indent The highest magnetic field investigated by Lortz \emph{et al.}~\cite{Lortz2006September} with the shaking technique was $\mu_0H=6\, $T where they observed the largest latent heat for their assumed first-order phase transition. In Fig.~\ref{fig.PRB_2012_6T_100Hz_pdf} we present our data at $\mu_0H=6\, $T and a shaking-field frequency $f=100\, $Hz. Again, no superimposed peak can be discerned in our data which we plotted in Fig.~\ref{fig.PRB_2012_6T_100Hz_pdf}. The same holds true for our $10\, $Hz measurements presented in Fig.~\ref{fig.PRB_2012_6T_10Hz_pdf}. For the measurement with $h_{ac}=0.52\, $mT in Fig.~\ref{fig.PRB_2012_6T_10Hz_pdf}a, the self-heating already sets in at about $14.1\, $K and generates a broad dip structure in the data in the region where Lortz \emph{et al.}~\cite{Lortz2006September} observed a small sharp peak in the specific heat. However, no such peak is discernible in our data in Figs.~\ref{fig.PRB_2012_6T_10Hz_pdf}a or \ref{fig.PRB_2012_6T_10Hz_pdf}b. As we further increase the shaking amplitude $h_{ac}$ at a fixed frequency $f=10\, $Hz in Fig.~\ref{fig.PRB_2012_6T_10Hz_both_pdf}, a broad dome-like feature is "carved out" by the shaking field. This dome-like feature seems to be superimposed to an extended self-heating dip. We drew a pink dashed line into Fig.~\ref{fig.PRB_2012_6T_10Hz_both_pdf}b in order to sketch the suspected shape of the self-heating dip how it would appear without the dome-like feature. The self-heating dip has expanded on the temperature scale to lower temperatures and its depth increased compared to the orange curve due to the increased shaking amplitude. According to the collective pinning theory by Larkin and Ovchinnikov \cite{Larkin1979}, the pinning strength decreases with increasing temperature (the peak effect is an exception). Decreasing the temperature at a constant magnetic field increases the pinning strength. A stronger shaking amplitude can overcome the stronger pinning at lower temperatures and make the vortices also \emph{walk} at lower temperatures leading to the observed expansion of the self-heating dip towards lower temperatures for higher $h_{ac}$. We marked in Fig.~\ref{fig.PRB_2012_6T_10Hz_both_pdf}b the onset of the self-heating dip as $T_{sdo}$ (\emph{sdo} stands for self-heating-dip onset) and the first sharp step-like appearance of self heating on increasing the temperature as $T_{so}$ (\emph{so} stands for self-heating onset). The maximum at the dome-like feature, which we identified with the center of the peak-effect region, we marked with $T_p$. As it appears, there are two different regions of self heating. On heating up during the measurement, the first sign of self heating sets in at $T_{so}$, where the curve suddenly drops very sharply about $5\, $mJ gat$^{-1}$ K$^{-2}$ below the shaking free $C/T$-curve and runs from $T_{so}$ on more or less parallel to the shaking free $C/T$-curve until it reaches the second region of self heating where at $T_{sdo}$ the self-heating dip sets in. With increasing shaking amplitude $h_{ac}$, the sharp onset of the first sign of self heating $T_{so}$ is shifted to lower temperatures in line with the collective pinning theory where for lower temperatures the pinning strength is stronger. In the regions where self heating is present, the magnetic vortices continuously enter the sample on one side of the sample, "walk" through the sample and leave it on the opposite side of the sample, thereby continuously generating heat inside the sample.\\
\begin{figure}
\includegraphics[width=75mm,totalheight=200mm,keepaspectratio]{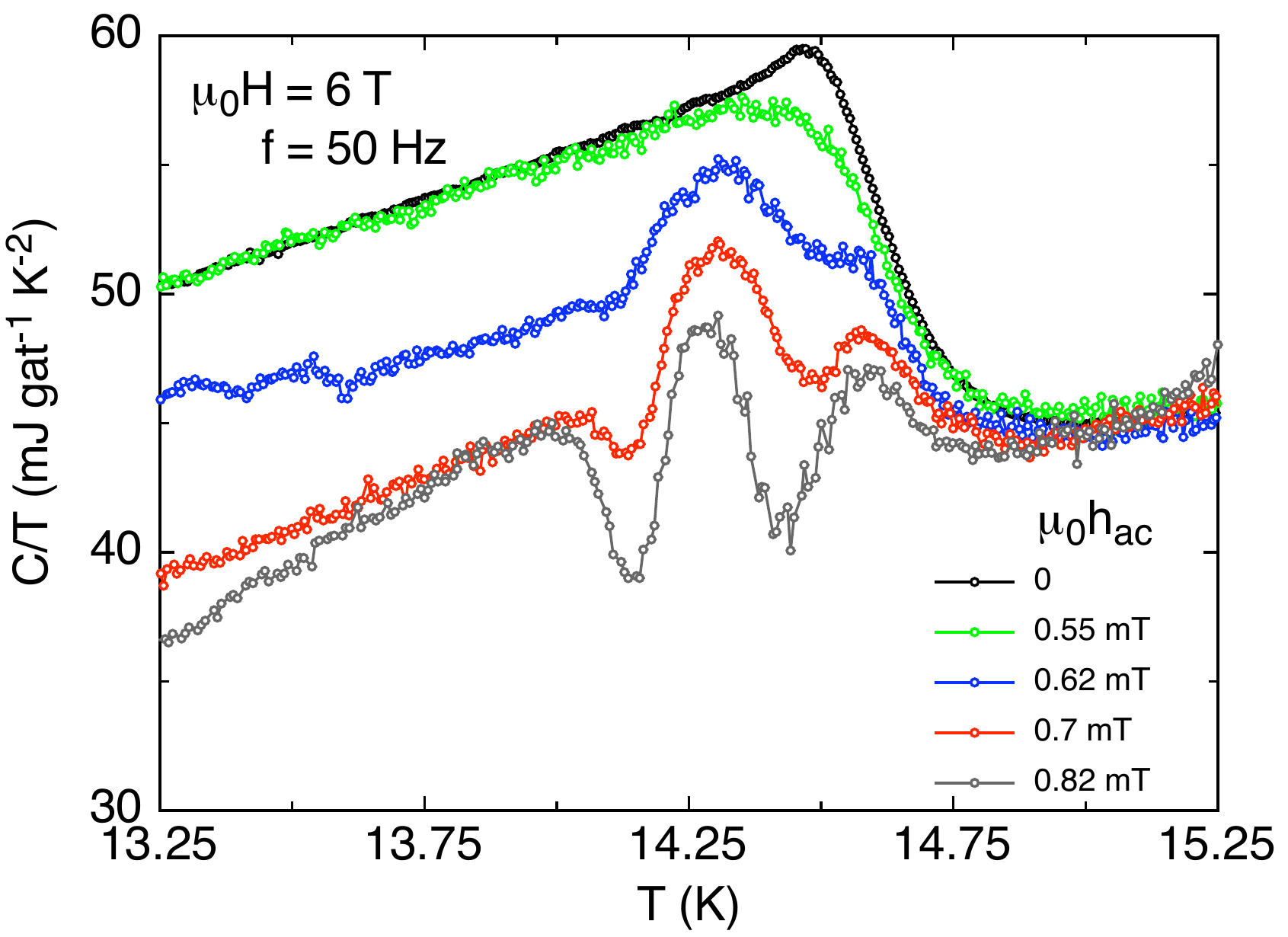}
\caption{Distorted specific heat of Nb$_3$Sn at $\mu_0H=6\, $T. A small shaking field with frequency $f=50\, $Hz was applied perpendicular to the main magnetic field $H$ at different amplitudes $h_{ac}$. The FC process prior to the measurement was very slow.\label{fig.PRB_2012_6T_50Hz_pdf}}
\end{figure}
In Fig.~\ref{fig.PRB_2012_6T_50Hz_pdf} we show data at $\mu_0H=6\, T$ and $f=50\, $Hz for different shaking amplitudes $h_{ac}$. We zoomed in on the peak-effect region near $T_c$, the sharp step-like onset of the self heating takes place at lower temperatures outside the zoomed in region. Again, a dome-like feature is "carved out" by the shaking field at the peak-effect region which has lower self heating due to the increased pinning strength in this region. This dome-like feature is broader than the peak-like feature observed by Lortz \emph{et al.}~\cite{Lortz2006September} at their supposed melting transition. In addition, the temperature of the center of the dome seems to be lower than the center of the supposed melting transition observed by Lortz \emph{et al.}~\cite{Lortz2006September}. Therefore it appears not very likely that the reduced self heating at the peak effect might have led to the peak observed by Lortz \emph{et al.}~\cite{Lortz2006September} at their supposed melting transition. However, one has to keep in mind, that different specific-heat methods and different shaking methods were used by the two groups. Interestingly, in Fig.~\ref{fig.PRB_2012_6T_50Hz_pdf} for higher amplitudes the dome-like feature is embedded in regions of increased self heating which appear as dips. According to an explanation of the peak effect by Pippard \cite{Pippard1969}, the vortex lattice softens in the peak-effect region, i.e., the shear modulus $C_{66}$ is vastly reduced, the vortex lattice becomes less rigid, and the vortices can bend and adjust better to the pinning sites. One may interpret our data in the way that the onset of the self-heating dip at $T_{sod}$ is the onset of the reduction of the shear modulus and the resulting less rigid vortex lattice is more prone to accelerations by the shaking field thereby leading to more self heating which causes the observed self-heating dip. It is known that the presence of pinning causes a friction during the motion of the vortices which leads to the heat dissipation. The bending of the vortices is crucial in this mechanism. When the shear modulus $C_{66}$ softens, the vortices can bend more easily due to the alternating forces exerted by the shaking field. On further increasing the temperature the shear modulus continues to reduce and the mechanism of better adjustment to the pinning sites sets in which increases the pinning strength and thereby hinders vortex motion in the region of the peak at $T_p$, at least in a part of the sample. The dome-like feature at $T_p$ inside the peak-effect region, superimposed to the self-heating dip, is in this picture the consequence of the increased pinning strength in the peak-effect region, since a hindered vortex motion also reduces the self heating.\\
\begin{figure}
\includegraphics[width=75mm,totalheight=200mm,keepaspectratio]{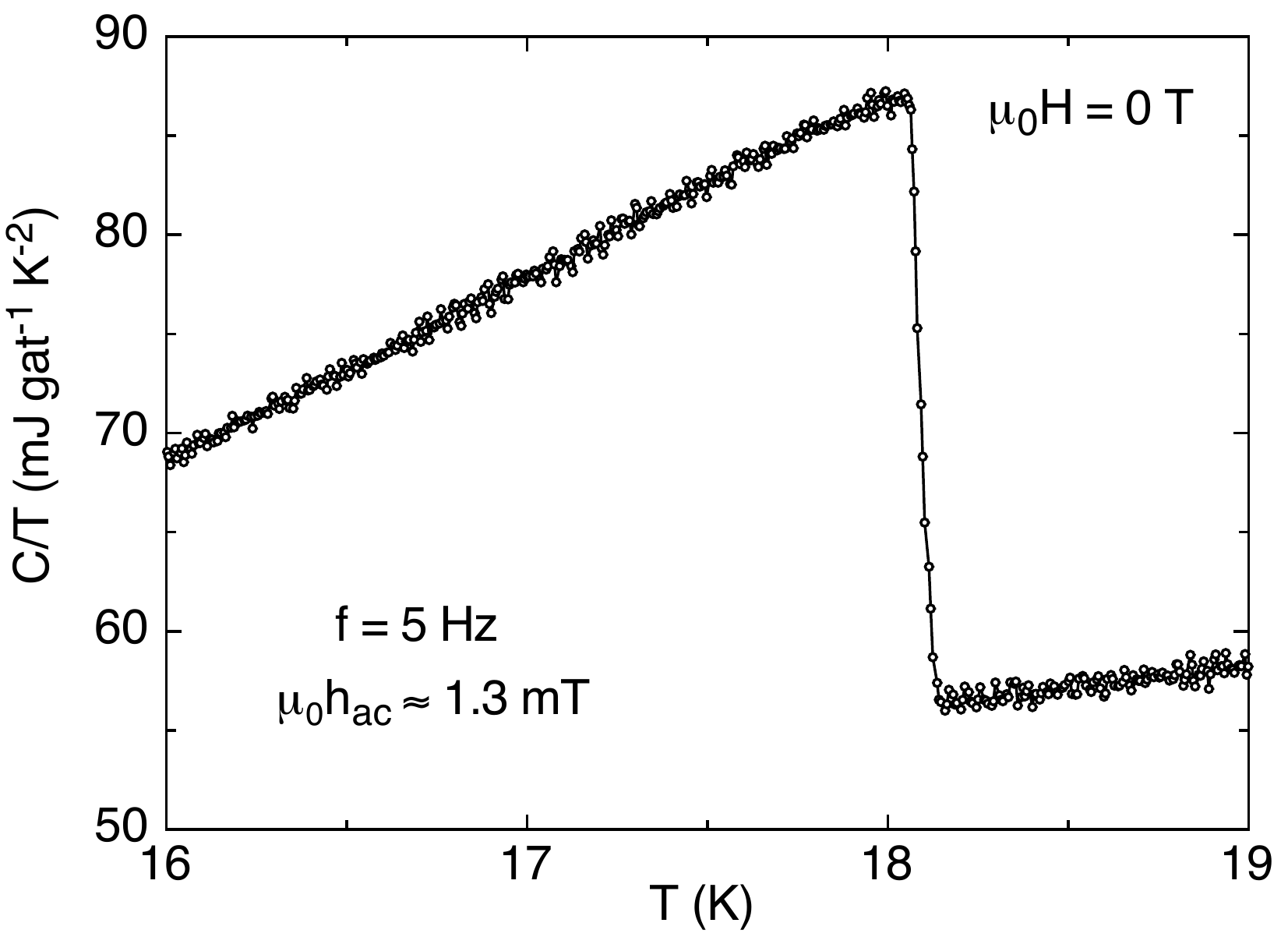}
\caption{Specific heat of Nb$_3$Sn at $\mu_0H=0\, $T. A small shaking field with $h_{ac} \approx 1.3\, $mT and $f=5\, $Hz was applied perpendicular to the main magnetic field $H$.\label{fig.PRB_2012_0T_5Hz_pdf}}
\end{figure}
\indent We next want to decide the question wether the self-heating dip is really caused by the motion of vortices or caused by different means like, e.g., the increase of quasiparticles with increasing temperature in the picture of the two-fluid model, where eddy currents made up of quasiparticles may cause the self heating. If the self-heating dip is mainly caused by vortex motion, it should disappear or at least diminish strongly in zero magnetic field despite the presence of a shaking field. We present in Fig.~\ref{fig.PRB_2012_0T_5Hz_pdf} a measurement at $\mu_0 H = 0\, $T; the shaking field had the frequency $f = 5\, $Hz and the rather high amplitude $h_{ac} \approx 1.3\, $mT. No sign of self-heating can be observed near $T_c$. We therefore conclude that the above observed dip in our data close to $T_c$ for $\mu_0H>3\, $T is most likely caused by self heating due to the motion of magnetic vortices.\\
\begin{figure}
\includegraphics[width=75mm,totalheight=200mm,keepaspectratio]{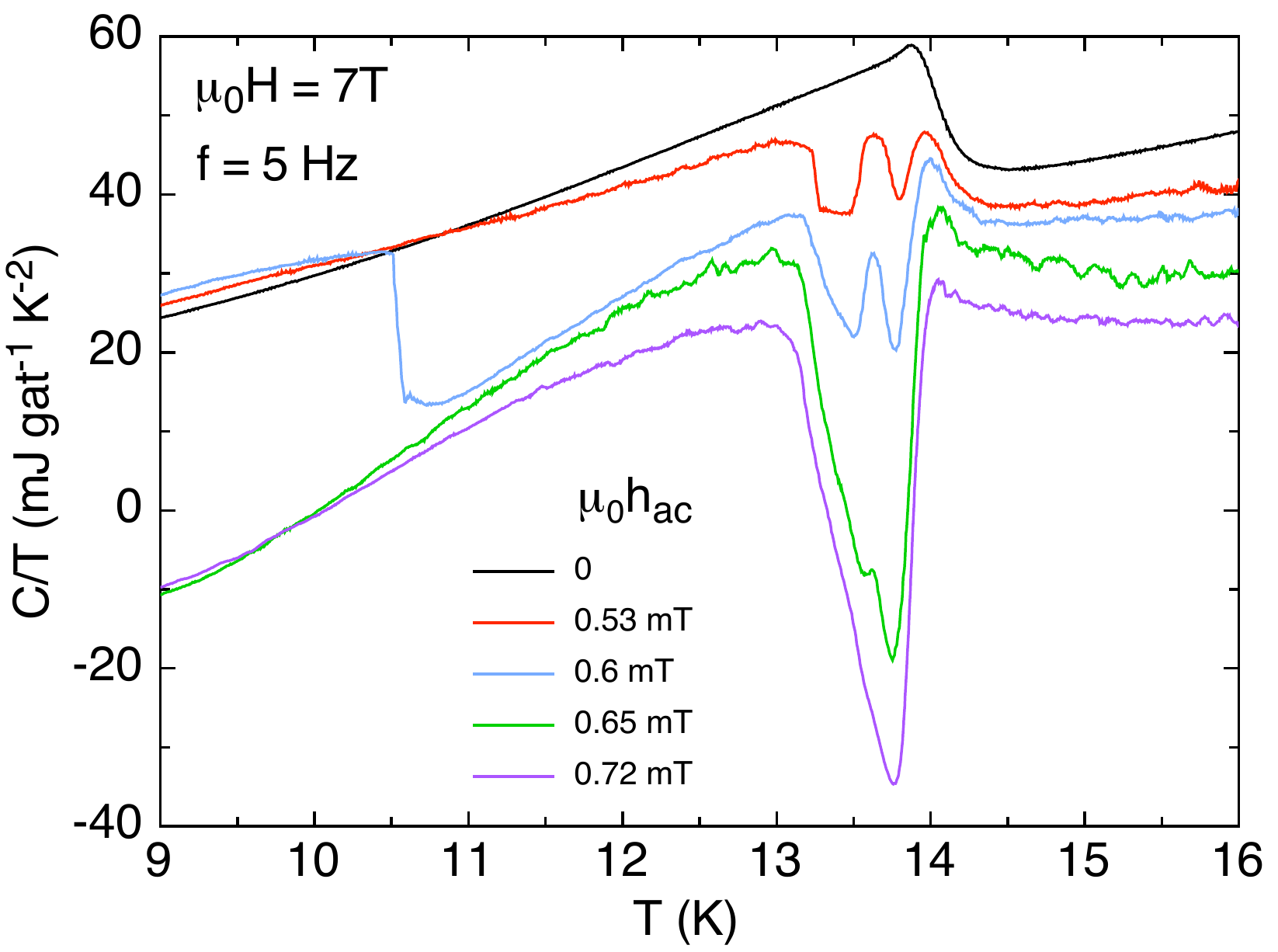}
\caption{Distorted specific heat of Nb$_3$Sn at $\mu_0H=7\, $T. A small shaking field with frequency $f=5\, $Hz was applied perpendicular to the main magnetic field $H$ at different amplitudes $h_{ac}$. The data were shifted for clarity.\label{fig.PRB_2012_7T_5Hz_pdf}}
\end{figure}
\indent If our interpretation of the dome-like feature as a manifestation of the increased pinning strength in the peak-effect region is correct, then further increasing the shaking amplitude $h_{ac}$ should be able to overcome the pinning strength even in the peak-effect region. Increasing the shaking amplitude was not possible for us due to the limitations of our shaking coil. However, we were able to increase the main magnetic field to $\mu_0H=7\, $T. The effectiveness of the shaking field increases with increasing main magnetic field, because the pinning strength decreases with increasing magnetic field as can be seen from the hysteresis in the magnetization loops of Nb$_3$Sn (see Refs.~\cite{Lortz2007March,Reibelt2010March} for magnetization measurements of this Nb$_3$Sn sample). In Fig.~\ref{fig.PRB_2012_7T_5Hz_pdf} we show data at $\mu_0H=7\, T$ and $f=5\, $Hz for different shaking amplitudes $h_{ac}$, the data was shifted for clarity. For medium shaking-field amplitudes $h_{ac}$ again the dome-like feature embedded in a self-heating dip emerges. Strikingly, for a large shaking-field amplitude $h_{ac}=0.65\, $mT (green curve) the dome-like feature diminishes and for the largest amplitude $h_{ac}=0.72\, $mT (purple curve) the dome-like feature even vanishes completely. If the dome-like feature origins from a first-order vortex-lattice melting transition, this vanishing for high shaking-filed amplitudes $h_{ac}$ would be unexpected. However, in the picture of our interpretation where the dome-like feature is a manifestation of the increased pinning strength in the peak-effect region, the vanishing of the dome-like feature with increasing $h_{ac}$ is expected.\\
\begin{figure}
\includegraphics[width=75mm,totalheight=200mm,keepaspectratio]{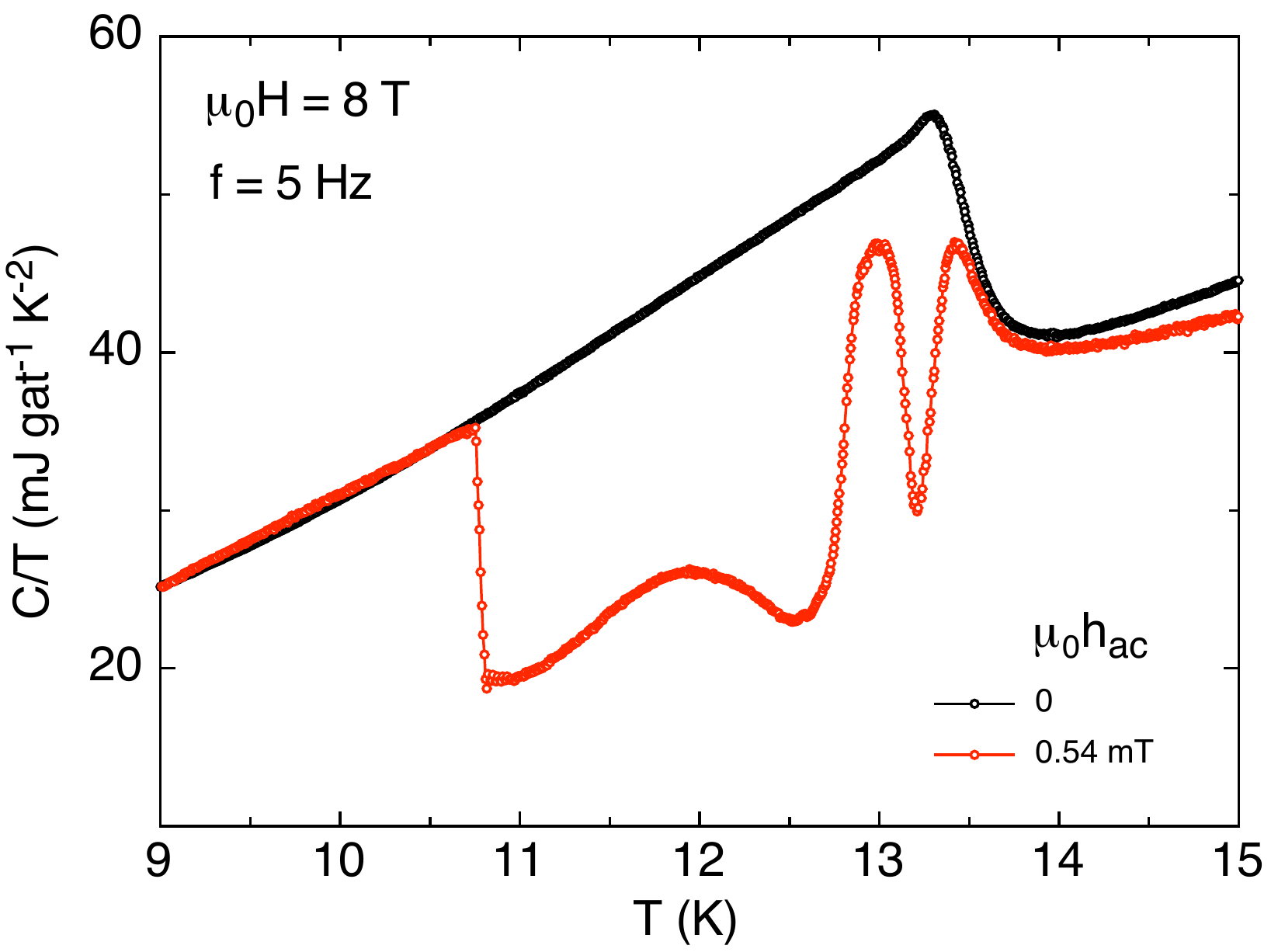}
\caption{Distorted specific heat of Nb$_3$Sn at $\mu_0H=8\, $T. A small shaking field with frequency $f=5\, $Hz and $h_{ac} = 0.54\, $mT was applied perpendicular to the main magnetic field $H$.\label{fig.PRB_2012_8T_5Hz_pdf}}
\end{figure}
\begin{figure}
\includegraphics[width=75mm,totalheight=200mm,keepaspectratio]{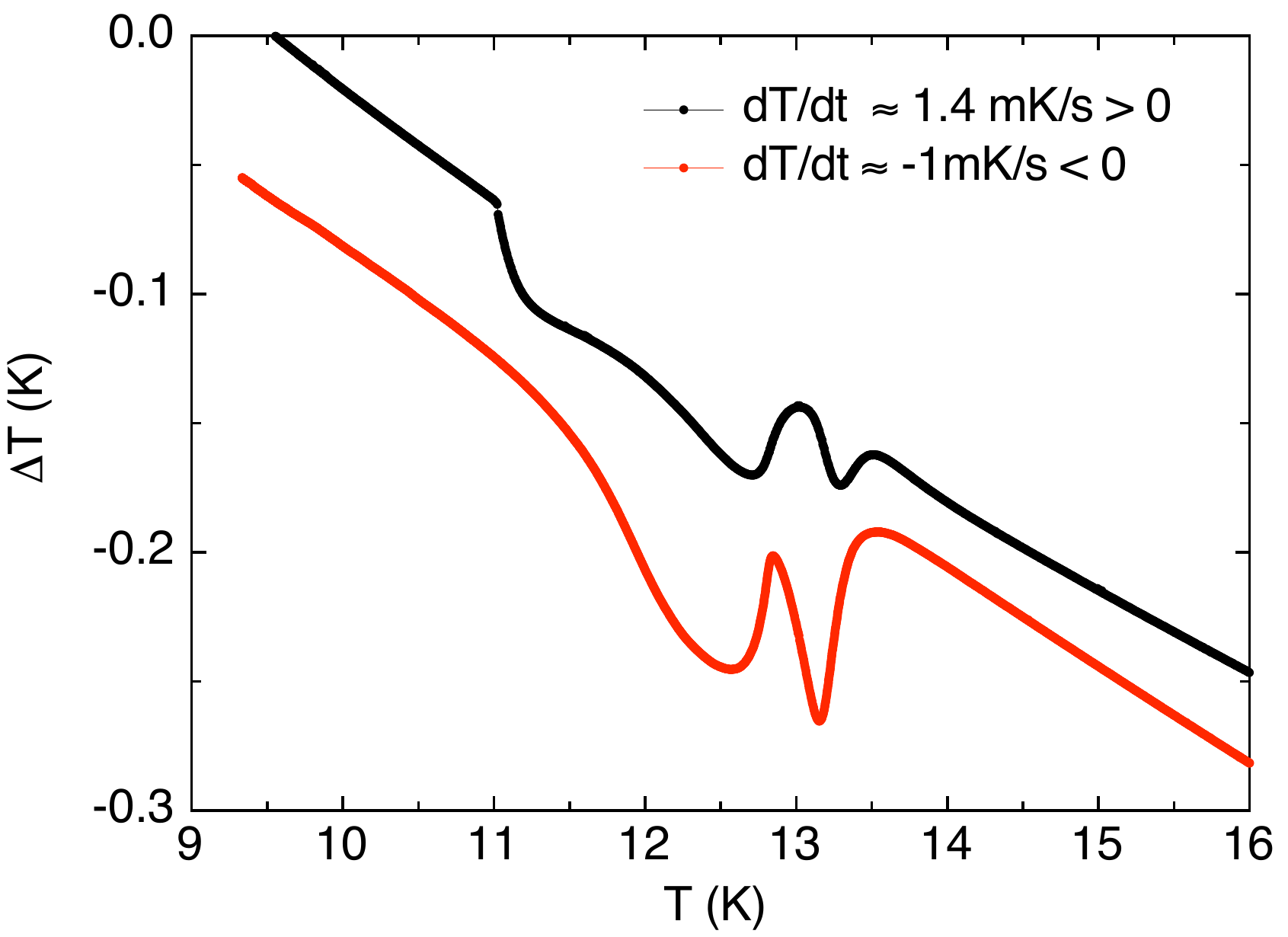}
\caption{Raw magnetocaloric data $\Delta T$ for increasing and decreasing temperature. The data were shifted for clarity.\label{fig.PRB_2012_up_down_comparison_pdf}}
\end{figure}
\indent We next present data at $\mu_0 H = 8\, $T. In Fig.~\ref{fig.PRB_2012_8T_5Hz_pdf} the typical self-heating pattern emerges again. In Fig.~\ref{fig.PRB_2012_up_down_comparison_pdf} we plotted the raw magnetocaloric data $\Delta T=T_r-T_s$, where $T_r$ is the temperature of the reference thermometer and $T_s$ is the temperature of the sample thermometer where the sample is mounted. We plotted data for a measurement on increasing the temperature and for a measurement on decreasing the temperature. Strikingly, a similar self-heating pattern emerges for both measurements in the peak-effect region. If the dome-like feature would have been due to a first-order melting transition, the related latent heat should have been released in one measurement and absorbed in the other measurement. However, we observe for both measurements a dome-like feature where the sample gets colder. The occurrence of self heating is rather cumbersome when it comes to the identification of phase transitions. The broad and huge shape of the dome-like feature already disqualify it as a manifestation of a melting transition to some degree. However, due to the comparison of a measurement on increasing with a measurement on decreasing the temperature, we were able to definitely exclude a first-order melting transition as the origin for the dome-like feature.\\
\begin{figure}
\includegraphics[width=75mm,totalheight=200mm,keepaspectratio]{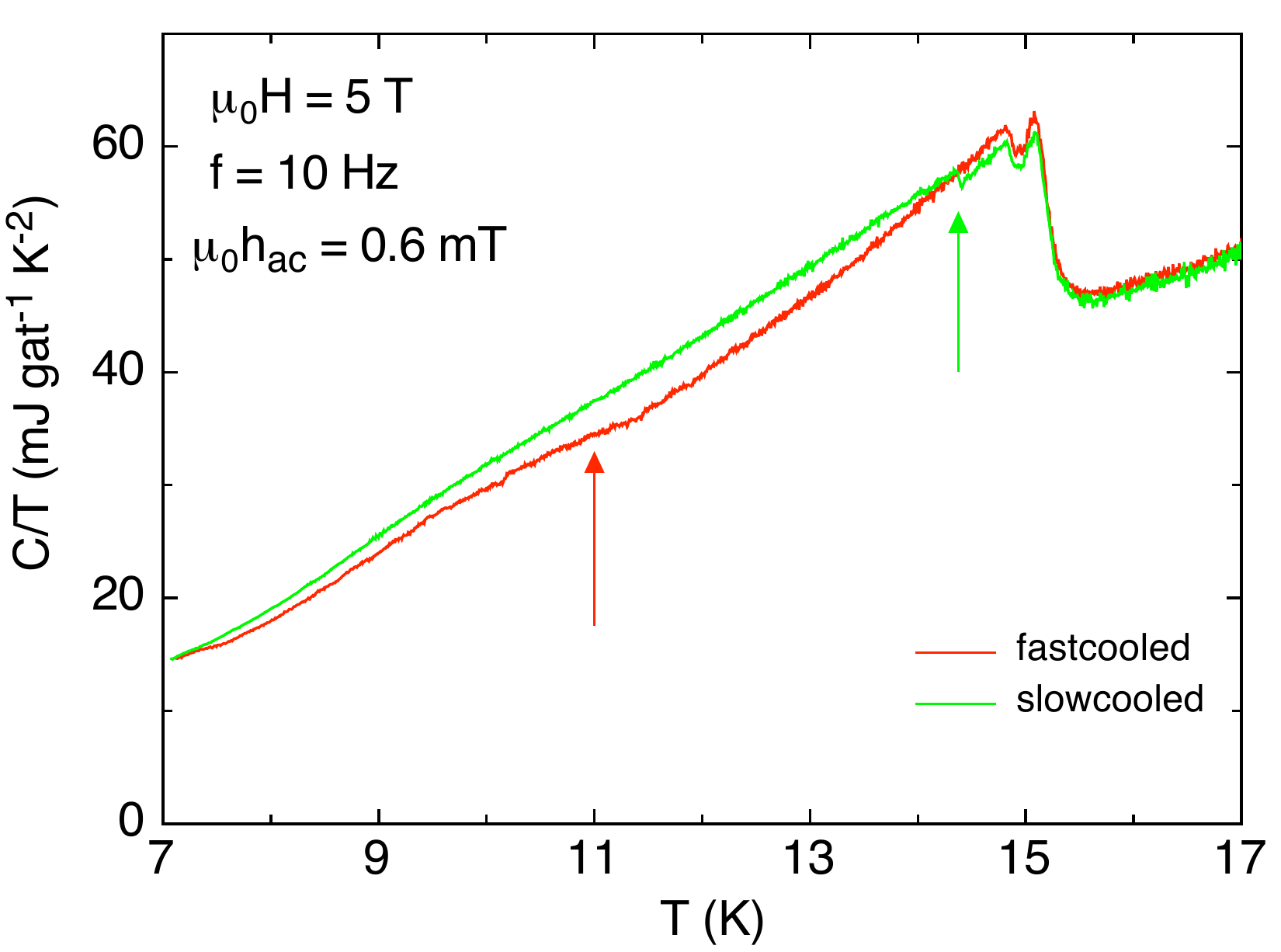}
\caption{Comparison of a measurement with slow cooled FC procedure preparation with a measurement with fast cooled FC procedure preparation.\label{fig.PRB_2012_comparison_slow_fast_pdf}}
\end{figure}
\indent Finally, we want to report on the influence of the cooling rate during the FC procedure on the self-heating pattern. In Fig.~\ref{fig.PRB_2012_comparison_slow_fast_pdf} we plotted two measurements at $\mu_0H=5\, $T, $f=10\, $Hz, and $\mu_0h_{ac}=0.6\, $mT. For the green curve we used the usual slow cooled FC procedure and the usual sharp step-like onset of the self heating occurs in the data, which we marked with a green arrow. For the red curve we used a very high cooling rate by the use of a cooling clamp which connected the measuring cell directly with the cold insert. It appears as if the formerly very sharp step-like onset of self heating turned into a smeared kink, which we marked with a red arrow in the figure. The faster cooling lead to a fast passing of the peak-effect region, giving the vortex lattice not enough time to adjust to the strongest pinning centers while the vortex lattice is easy bendable in the peak-effect region where the shear modulus is reduced. The ordered weakly pinned vortex lattice gets quenched, it has not enough time to pick up much disorder and strong pinning before it becomes rigid at low temperatures where it cannot adjust to the strongest pinning centers anymore. On heating up the sample again during the measurement, the less strong pinned vortex lattice allows the shaking field already at lower temperatures to cause vortex motion. The broadening of the onset may be caused by the circumstance that some parts of the sample are stronger pinned than others for the fast cooled FC procedure. In contrast, for the slow cooled FC procedure the sample stays for a long enough time in the peak-effect region on cooling the sample, allowing the vortices throughout the whole sample to adjust to the strongest pinning centers, which leads to a homogeneous strong pinning throughout the whole sample and as a result to the sharper self-heating onset at higher temperatures on shaking during the measurement.\\
\indent This finding is likely not a new one. However, we want to mention its practical relevance. It appears to be beneficial to cool down type-II superconductors very slowly across their peak-effect region (if present) in the presence of a magnetic field, in order to achieve (freeze in) the strongest pinned and most homogeneous phase throughout the whole superconductor.

\section{CONCLUSION}
To conclude, we confirm the observation of the fluctuation peak in the specific heat of Nb$_3$Sn near $T_c(H)$, which was previously reported by Lortz \emph{et al.}~\cite{Lortz2006September,Lortz2007April} for the same crystal. Along with work done by other authors our data support the view that thermal fluctuations manifest not only in high-$T_c$ superconductors but also in low-$T_c$ superconductors like Nb$_3$Sn \cite{Lortz2006September,Lortz2007April}, Nb \cite{Farrant1975}, and dirty Bi$_{0.4}$Sb$_{0.6}$ films \cite{Zally1971December}.\\
\indent However, we were not able to observe the by Lortz \emph{et al.}~\cite{Lortz2006September} reported melting of the vortex lattice in this Nb$_3$Sn crystal. We observed no small sharp peak at the by the measurements of Lortz \emph{et al.}~\cite{Lortz2006September} indicated position in the magnetic phase diagram in any of our measurements. We want to point out that our data at $\mu_0H=3\, $T and even some of our data at $\mu_0H=4\, $T are virtually free of self heating and still we do not observe any sign of a vortex melting. The by us observed broad dome-like feature in the measurements at higher magnetic fields where self heating is present, is interpreted by us as a manifestation of the increased pinning in the peak-effect region. We want to emphasize again that the discrepancy between the data of Lortz \emph{et al.}~\cite{Lortz2006September} and our data presented in this work may originate from the different specific-heat methods and different shaking methods and parameters used by the two groups. However, it would be interesting to repeat the measurements of Lortz \emph{et al.}~\cite{Lortz2006September} with their ac technique but with a truly transversal or longitudinal vortex-shaking configuration. In addition, higher shaking-field amplitudes may be used with their already existing setup in order to exclude a vanishing of the supposed vortex-lattice melting peak as it was the case for the dome-like feature in our data.

\section{ACKNOWLEDGMENT}
We thank the group of A.~Schilling for their support. This work was supported by the Schweizerische Nationalfonds zur F\"{o}rderung der Wissenschaftlichen Forschung, Grants No.~$20$-$111653$ and No.~$20$-$119793$.


\end{document}